
%
%
%
\input harvmac.tex
\def\dW{\Delta W}
\let\bmx=\bordermatrix
\let\bar=\overline
\def\nl{\hfill\break}
\let\<=\langle
\let\>=\rangle
\let\e=\epsilon
\let\tht=\theta
\let\t=\theta

\let\l=\lambda
\let\p=\prime
\def\Sn{S^{N=2}[n](\t )}
\def\Zn{${\bf Z}_n$}
\noblackbox
\pretolerance=750
\lref\LGSol{P. Fendley, S.D. Mathur, C. Vafa and N.P. Warner,
Phys. Lett. B243 (1990) 257}
\lref\pk{P. Fendley and K. Intriligator, {\it Scattering and
Thermodynamics of Fractionally-Charged Supersymmetric Solitons}, BUHEP-91-17,
HUTP-91/A043, to appear in Nucl. Phys. B}
\lref\rJR{R. Jackiw and C. Rebbi, Phys. Rev. D13 (1976) 3398}
\lref\rGW{J. Goldstone and F. Wilczek, Phys. Rev. Lett. 47 (1981) 986}
\lref\rHKSS{W.P. Su and J.R. Schrieffer, Phys. Rev. Lett. 46 (1981) 738;\nl
A.J. Heeger, S. Kivelson, J.R. Schrieffer, and W.P. Su, Rev. Mod.
Phys., 60 (1988) 781}
\lref\rZamint{A.B. Zamolodchikov, JETP Lett. 46 (1987) 160}
\lref\rVW{A.B. Zamolodchikov Sov. J. Nucl. Phys. 44 (1986) 529;\nl
D. Kastor, E. Martinec and S. Shenker, Nucl. Phys. B316 (1989) 590;\nl
E. Martinec, Phys. Lett. 217B (1989) 431;\nl
C. Vafa and N.P. Warner, Phys. Lett 218B (1989) 51}
\lref\rWO{E. Witten and D. Olive, Phys. Lett. 78B (1978) 97}
\lref\toda{P. Fendley, W. Lerche, S.D. Mathur and N.P. Warner,
Nucl. Phys. B348 (1991) 66}
\lref\rHollo{T. Hollowood, ``Quantum Solitons in Affine Toda
Field Theories'', Princeton preprint PUPT-1286, hepth\#9110010}
\lref\rNS{A.J. Niemi and G.W. Semenoff, Phys. Rep. 135 (1986) 99}
\lref\sf{A. Schwimmer and N. Seiberg, Phys. Lett. B184 (1987) 191}
\lref\rZandZ{A.B. Zamolodchikov and Al.B. Zamolodchikov, Ann.
Phys. 120 (1980) 253}
\lref\rAZi{A.B. Zamolodchikov, Adv. Stud. Pure Math. 19 (1989) 1}
\lref\rKS{R. K\"oberle and J.A. Sweica, Phys. Lett. 86B (1979) 209}
\lref\rKMi{T.R. Klassen and E. Melzer, Nucl. Phys. B338 (1990)
485; B350 (1990) 635}
\lref\rKuR{P.P. Kulish and N. Yu. Reshetikhin, Sov. Phys. JETP 53
(1981) 108; J. Phys. A16 (1983) L591}
\lref\rKK{R. K\"oberle and V. Kurak, Phys. Lett. 191B (1987) 295}
\lref\rYY{C.N. Yang and C.P. Yang, J. Math. Phys. 10 (1969) 1115}
\lref\tbaref{Al.B. Zamolodchikov, Nucl. Phys. B342 (1990) 695}
\lref\rAlZii{Al.B. Zamolodchikov, Nucl. Phys. B358 (1991) 497}
\lref\rFST{E.K. Sklyanin, L.A. Takhtadzhyan and L.D. Faddeev,
Theo. Math. Phys. 40 (1980) 688. L. Faddeev, in Les Houches 1982 {\it Recent
advances in field theory and statistical mechanics} (North-Holland 1984),
edited by J.-B.  Zuber and R. Stora}
\lref\rBL{D. Bernard and A. LeClair, Phys. Lett. 247B (1990) 309}
\lref\rBCN{H.W.J. Bl\"ote, J.L. Cardy and M.P. Nightingale, Phys.
Rev.  Lett. 56 (1986)742;\nl I. Affleck, Phys. Rev. Lett., 56 (1986) 746}
\lref\rKR{A.N. Kirillov and N. Yu. Reshetikhin, J. Phys. A20
(1987) 1587}
\lref\rLewin{L. Lewin, {\it Polylogarithms and Associated
Functions} (North-Holland, Amsterdam, 1981)}
\lref\rGep{D. Gepner, {\it Fusion Rings and Geometry} preprint
NSF-ITP-90-184}
\lref\lwv{W. Lerche, C. Vafa, N. Warner, Nucl.  Phys. B324 (1989)
427}
\lref\dvv{R. Dijkgraaf, E. Verlinde, and H. Verlinde, Nucl. Phys.
B352 (1991) 59}
\lref\rken{K. Intriligator, Mod. Phys. Lett. A6 (1991) 3543}
\lref\rABL{C. Ahn, D. Bernard and A. LeClair, Nucl. Phys. B346
(1990) 409}
\lref\rLW{W. Lerche and N.P. Warner, Nucl. Phys. B358 (1991) 571}
\lref\rCV{S. Cecotti and C. Vafa, Nucl. Phys. B367 (1991) 359}
\lref\bonus{K. Intriligator, Nucl. Phys. B332 (1990) 541}
\lref\rNar{K.S. Narain, Nucl. Phys. B243 (1984) 131}
\lref\witN{E. Witten, Nucl. Phys. B149 (1979) 285}
\lref\witten{E. Witten, Nucl. Phys. B340 (1990) 281}
\lref\qrings{C. Vafa, {\it Topological Mirrors and Quantum
Rings}, Harvard preprint HUTP-91/A059}
\lref\intcpnrefs{E. Abdalla, M.C. Abdalla, and M. Gomes, Phys.
Rev. D25 (1982) 425; Phys. Rev. D27 (1983) 825}
\lref\rCVii{S. Cecotti and C. Vafa, {\it Exact Results for
Supersymmetric Sigma Models}, Harvard preprint HUTP-91/A062}
\lref\cpnrefs{E. Abdalla and A. Lima-Santos, Phys. Rev. D29
(1984) 1851;\nl
R. K\"oberle and V. Kurak, Phys. Rev. D36 (1987) 627}
\lref\grrefs{E. Abdalla, M. Forger, and A. Lima-Santos, Nucl. Phys. B256
(1985) 145; E. Abdalla and A. Lima-Santos, Phys. Lett. B206 (1988) 281}
\lref\rJKMO{M. Jimbo, A. Kuniba, T. Miwa and M. Okado, Comm.
Math. Phys. 119 (1988) 543}
\lref\RSOS{A.B. Zamolodchikov, Landau Institute preprint,
September 1989}
\lref\rABF{G.E. Andrews, R.J. Baxter and P.J. Forrester, J. Stat.
Phys. 35 (1984) 193}
\lref\rBR{V. Bazhanov and N. Reshetikhin, J. Phys. A23 (1990)
1477}
\lref\newZ{A.B. Zamolodchikov and Al. B. Zamolodchikov,
{\it Massless Factorized Scattering and Sigma Models with
Topological Terms}, Ecole Normale preprint}
\lref\un{P. Fendley, UnTexed}
\lref\Kob{K. Kobayashi and T. Uematsu, {\it Quantum Conserved
Charges and S-matrices in $N$=2 Supersymmetric Sine-Gordon Theory} KUCP-41
(1991), hepth\#9112043}
\lref\rKzS{Y. Kazama and H. Suzuki, Phys. Lett. 216B (1989) 112;
Nucl. Phys. B321 (1989) 232}
\lref\rEYHM{T. Eguchi and S.-K. Yang, Phys. Lett. 224B (1989)
373;\nl T. Hollowood and P. Mansfield, Phys. Lett. 226B (1989) 73}
\lref\rAlZiv{Al.B. Zamolodchikov, {\it Resonance Factorized
Scattering and Roaming Trajectories}, Ecole Normale preprint ENS-LPS-355}
\lref\rTsv{A.M. Tsvelik, Sov. J. Nucl. Phys. 47 (1988) 172}
\lref\rTS{M. Takahashi and M. Suzuki, Prog. Th. Phys. 48 (1972) 2187}
\lref\rAZtci{M.~L\"assig, G.~Mussardo and J.L.~Cardy, Nucl.~Phys.~B348 (1991)
591; \nl
F.A.~Smirnov, Int.~J.~Mod.~Phys.~A6 (1991) 1407;\nl
A.B. Zamolodchikov, {\it S-matrix of the Subleading Magnetic
Perturbation of the Tricritical Ising Model}, Princeton preprint
PUPT-1195-90;\nl
F. Colomo, A. Koubek and G.  Mussardo, {\it On the S-Matrix
of the Sub-leading Magnetic Deformation of the Tricritical Ising Model in Two
Dimensions}, ISAS preprint 94/91/EP}
\lref\rBLii{N. Yu. Reshetikhin and F. Smirnov, Comm. Math.
Phys 131 (1990) 157;\nl
D. Bernard and A. LeClair, Nucl. Phys. B 340 (1990) 721. }
\lref\rKar{M. Karowski, Nucl. Phys. B300 (1988) 473}
\lref\rBel{A.A. Belavin, Nucl. Phys. B180 (1980) 189}
\lref\rGin{P. Ginsparg, Nucl. Phys. B295 (1988) 153}
\lref\rpara{A.M. Tsvelik, Nucl. Phys. B305 (1988) 675;\nl
V.A. Fateev and A.B. Zamolodchikov, Int. J. Mod. Phys. A5 (1990) 1025}
\lref\rMW{P. Mathieu and M. Walton, Phys. Lett. 254B (1991) 106.}

\Title{\vbox{\baselineskip12pt\hbox{BUHEP-92-5, HUTP-91/A067,
hepth@xxx/9202011}}}
{\vbox{\centerline{Scattering and Thermodynamics in Integrable
$N$=2
Theories}}}
\centerline{P. Fendley}
\bigskip\centerline{Department of Physics}
\centerline{Boston University}\centerline{590 Commonwealth
Avenue}
\centerline{Boston, MA 02215}
\bigskip\centerline{K. Intriligator}
\bigskip\centerline{Lyman Laboratory of Physics}
\centerline{Harvard University}\centerline{Cambridge, MA 02138}
\vskip .2in

We study $N$=2 supersymmetric integrable theories with
spontaneously-broken \Zn\ symmetry.  They have exact soliton masses given by
the affine $SU(n)$ Toda masses and fractional fermion numbers given by
multiples of $1/n$. The basic such $N$=2 integrable theory is the
$A_n$-type $N$=2 minimal model perturbed by the most relevant operator. The
soliton content and exact S-matrices are obtained using the Landau-Ginzburg
description.  We study the thermodynamics of these theories and calculate the
ground-state energies exactly, verifying that they have the correct conformal
limits.  We conjecture that the soliton content and S-matrices in other
integrable \Zn\ $N$=2 theories are given by the tensor product of the above
basic $N$=2 \Zn\ scattering theory with various $N$=0 theories. In particular,
we consider integrable perturbations of $N$=2 Kazama-Suzuki models described
by generalized Chebyshev potentials, $CP^{n-1}$ sigma models, and $N$=2
sine-Gordon and its affine Toda generalizations.
\vskip .25in

\Date{1/92}


\newsec{Introduction}

Two-dimensional models with $N$=2 supersymmetry have been studied for a
variety of reasons.  Their interesting formal properties include their
important role in string theory and topological field theories, and the fact
that nonrenormalization theorems make many properties exactly calculable.  One
nice result of the nonrenormalization theorems is that the Landau-Ginzburg
description is very powerful: many properties of an $N$=2 theory follow solely
{}from its Landau-Ginzburg potential, without requiring knowledge of the
kinetic term. In particular, the soliton structure of an off-critical $N$=2
theory is easily obtained via the Landau-Ginzburg description \refs{\LGSol ,
\pk}.  With knowledge of this soliton structure in an integrable theory, the
exact S-matrix can often be calculated exactly.

An interesting fact also seen in the Landau-Ginzburg description is that the
solitons in $N$=2 theories generally have fractional fermion number.  It is
well-known that many 1+1 dimensional theories have fractional fermion number
in the soliton sectors \refs{\rJR,\rGW}. In fact, we do not know of any
integrable $N$=2 theories without fractional charges. Even theories such as
supersymmetric $CP^{n-1}$ sigma models, which naively appear not to have any
solitons at all, have solitons with fractional fermion number.  Fractional
fermion number is believed to occur in some $1+1$-dimensional polymer
systems \rHKSS\ (as well as in the $2+1$-dimensional fractional quantum Hall
effect!); our
exactly solvable $N$=2 systems may provide a way of further understanding
these theories.

We will show that the basic supersymmetric structure in $N$=2 integrable
theories having a spontaneously-broken \Zn\ symmetry is that of the $A_n$ type
$N$=2 minimal model perturbed by the most-relevant operator.  The solitons in
these theories are supersymmetric doublets with the affine $SU(n)$ Toda masses
and fractional charges given by multiples of $1/n$ (sect.\ 2).  We find (in
sect.\ 3) the exact S-matrices $\Sn$ for these basic theories.  In sect.\ 4 we
obtain integral equations for the exact ground-state (Casimir) energy $E(R)$
of the quantum field theories in a periodic box of size $R$, and verify that
they give the correct answers in the conformal and infra-red limits. This is
done by using the thermodynamic Bethe ansatz, and in the course of this
calculation we find a new kind of TBA system.

The scattering theory of essentially every other known integrable $N$=2
theory exhibits a
\Zn\ structure and is either shown or conjectured to be a tensor product of
the above basic \Zn-type $N$=2 integrable scattering theory with various
$N$=0 scattering theories:
\eqn\sprod{S^{N=2}(\t )=S^{N=0}(\t )\otimes \Sn .}
For example, in an earlier paper \pk, we discussed the $N$=2 minimal conformal
field theories models perturbed by their least relevant operator.  The
Landau-Ginzburg superpotentials are given by Chebyshev polynomials, which have
a ${\bf Z}_2$ structure; the solitons all have equal mass and fractional
charge $\pm\half$.  We showed using the $N$=2 algebra and verified using the
TBA that these scattering theories have such a tensor-product decomposition.
The soliton structure and exact S-matrices of these models were found to be
described by tensor products of $S^{N=2}[n=2](\tht)$ with the RSOS scattering
theories \RSOS\ describing the $N$=0 minimal models perturbed by the least
relevant operator.  This confirmed the results of \rBL, where this
tensor-product structure was conjectured because it was the simplest
structure consistent with the quantum-group symmetry of the models. We showed
in \pk\ that the perturbed $D_k$ and $E$ type $N$=2 theories also had this
same structure, with the $N$=0 part the RSOS theory for the appropriate Dynkin
diagram.  We find here that \sprod\ generalizes to other integrable $N$=2
theories.

While we know that an $N$=2 integrable theory with a \Zn\ symmetry will have
$\Sn$ for the $N$=2 part in \sprod, we emphasize that the remaining $N$=0 part
in \sprod\ is not always the S-matrix for the analogous $N$=0 theory.  The
S-matrix for supersymmetric $CP^{n-1}$, for example, is of the form \sprod,
but there is no immediately obvious guess for which $N$=0 theory enters in
\sprod, because the $N$=0 $CP^{n-1}$ sigma model is not even
integrable!

In sect.\ 5 we consider perturbations of more general Kazama-Suzuki $N$=2
theories, which are conjectured to be integrable.  The Landau-Ginzburg
superpotentials for these theories are the $k$-th generalized Chebyshev
polynomials in $n-1$ variables.  We conjecture that the soliton content and
S-matrices for these theories are tensor products \sprod\ of the basic $N$=2
part $\Sn$ with an $N$=0 part corresponding to an analogous integrable
perturbation of an $N$=0 $SU(n)$ coset theory.

In sect. 6 we consider the supersymmetric sigma model on $CP^{n-1}$.  Here
again, the soliton content and S-matrices (originally found in \cpnrefs) are a
tensor product \sprod\ of the $n$-th $N$=2 minimal integrable theory $\Sn$
with an $N$=0 structure.  In fact, the theories are described by the
$k\rightarrow \infty$ limit of the generalized Chebyshev theories discussed in
the last paragraph.  We construct the TBA system of integral equations
describing the $CP^1$ case.

In sect. 7 we discuss $N$=2 super sine-Gordon theory and its generalizations.
We relate these theories at their critical coupling to the $CP^{n-1}$ theories
by relating their S-matrices.  For arbitrary coupling we expect the theories
to be a tensor product with an S-matrix of the form \sprod .  For example,
$N$=2 sine-Gordon is of the form \sprod\ involving the $n$=2 case of the basic
$N$=2 theory and the S-matrix for $N$=0 sine-Gordon \Kob.  We provide the TBA
system and thereby verify this S-matrix.

\newsec{The Basic $N$=2 Integrable Theories}

We consider the ($A_n$) $N$=2 minimal conformal field theories perturbed by
the most-relevant operator preserving the $N$=2 supersymmetry.  These theories
are integrable; non-trivial conserved charges were constructed in the
perturbed conformal field theories in \LGSol, in the manner of \rZamint.
These charges are believed to exist at all integer spins except multiples of
$n$.  This perturbed theory can be described by the Landau-Ginzburg
superpotential \rVW
\eqn\mrp{W={X^{n+1}\over n+1}-\beta X .}
The Landau-Ginzburg description gives us a nice way of finding the soliton
structure \LGSol. The bosonic part of the superpotential is given by $|\del
W/\del X|^2$, so the vacua are given by the equation $X^{n}=\beta$. Scaling
$\beta=1$, they are the roots of unity $$X^{(j)}=\exp(j2\pi i/n),$$ forming a
polygon with $n$ vertices.  There is a fundamental soliton connecting any two
of the vacua.  We label the soliton which connects vacuum $X^{(j)}$ with
vacuum $X^{(j+r)}$, for $r=1,\dots ,n-1$, by $K_{j(j+r)}$ ($K$ for ``kink'').

Every soliton forms a multiplet under supersymmetry. It turns out
that there
is only one fermionic
zero mode in the presence of a soliton, so each soliton
$K_{j(j+r)}$ is a
doublet, which we label by
$(u_r,d_r)$, under the $N$=2 supersymmetry; the label $j$ of the
initial
vacuum has does not label distinct solitons because of the \Zn\
symmetry of
the vacua under $X\rightarrow \omega X$ with $\omega ^n=1$. The
action of the
$N$=2 generators on the $(u_r,d_r)$ solitons is as discussed in
\pk .  A
two-dimensional representation of the $N$=2 supersymmetry must
have mass saturating the Bogomolny bound: $m_r=|W(X^{(j+r)}) -
W(X^{(r)})|
\equiv |\dW|=M\sin (\pi r/n)$, where $M$=$2n/n+1$ \rWO; the
soliton mass is the length of the line connecting the appropriate vertices of
the polygon.  Note that the masses are exactly those of affine $SU(n)$ Toda
theory. Our theory is in fact an affine Toda theory with imaginary coupling
and a background charge \toda. There is no obvious reason why the mass
spectrum of the solitons in a theory with imaginary coupling is identical to
that of the particles in the same theory with real coupling, but in Toda
theories this is true classically and is stable to small quantum fluctuations
\rHollo.

The fermion number of the solitons also follows from the
Landau-Ginzburg
description.  Using standard arguments \refs{\rGW\ ,\rNS}
it is straightforward to derive that the fermion number $f_r$ of
$u_r$ is
\eqn\fnum{\eqalign{f_r&=-(2\pi)^{-1}\Delta Im \log W''\cr
&={r\over n}.\cr}}
A simple argument for this result is as follows.  In the soliton sector the
fermion number is
\eqn\ff{f=\int {dx\over 2\pi}(-i)(\partial -\bar\partial)\phi
_{sol}=-{1\over
2\pi}\Delta\phi _{sol},}
where the integrand is the bosonized fermion current $J-\bar J$ along the
classical soliton solution. The field $\phi _{sol}$ is normalized so that the
supersymmetry generators have charges $\pm 1$.  The chiral primary field
corresponding to one unit of left-right symmetric spectral flow
\sf\ is then represented by
\eqn\sf{{\cal U}_{1,1}= e^{i\phi _{sol}}.}
{}From eqns. \ff\ and \sf\ we have
\eqn\fff{f=-{1\over 2\pi}\Delta Im\log {\cal U}_{1,1}=-{1\over
2\pi }\Delta
Im\log \det (\partial _i\partial _jW),}
where we make use of the fact that the field ${\cal U}_{1,1}$ is proportional
to the Hessian determinant \lwv . In our one-chiral-superfield case, eqn.\fff\
reduces to \fnum .  Because $d_r$ is obtained from $u_r$ by the action of
$Q^-$ (or $\bar Q^+$), the $(u_r,d_r)$ doublets have charges $(f_r,
f_r-1)=(r/n,r/n-1)$ \pk.

Our TBA analysis will confirm that this simplest $N$=2 soliton structure is in
fact the correct soliton content for the perturbed theories
\mrp .  In \LGSol\ the supermultiplet was unnecessarily doubled
because proper account was not taken of the fractional charges. The TBA
analysis shows us that this is wrong: it leads to the wrong UV Casimir energy.

\newsec{The exact S-matrix}

Since our models are integrable, there are many constraints on the S-matrices
which allow us to derive them exactly \refs{\rZandZ,\rAZi}.  For example,
momenta are conserved individually (only quantum numbers can change in a
collision) and the $n$-body S-matrix factorizes into a product of two-body
ones. In addition, our S-matrices must commute with the $N$=2 supersymmetry
generators.  In this section, we will find the minimal scattering theory and
S-matrix consistent with these constraints.  There is the usual ambiguity of
adding extra particles corresponding to extra CDD poles in the S-matrix
prefactor.  The thermodynamic analysis of the next section verifies that our
minimal solution (with no extra CDD poles) gives the appropriate ultraviolet
limit, effectively confirming that it is the correct scattering theory.

The vacua boundary conditions at spatial infinity, kinematic
constraints, and
the conserved currents restrict the form of the S-matrix. When
scattering a
soliton of type $K_{j(j+r)}$ with rapidity $\t _1$ (i.e.
$u_{r}(\t _1)$ or
$d_{r}(\t _1)$) with a soliton of type $K_{(j+r)(j+r+s)}(\t _2)$,
the outgoing
solitons must be of the type $K_{j(j+s)}(\t _2)$ and
$K_{(j+s)(j+r+s)}(\t
_1)$.  In other words, when we scatter $u_r$ or $d_r$ with $u_s$
or $d_s$, the
$r$ and $s$ labels scatter diagonally: this is required for the
conserved
currents to be consistent with the bound-state structure, in the
same manner
as $A_n$ Toda theory \refs{\rKS, \rKMi}.  By fermion-number
conservation, the
S-matrix
for such scattering is of the form:
\eqn\Smat{\bmx{&d_su_r&u_sd_r\cr
u_rd_s&b_{r,s}(\t )&{\tilde c}_{r,s}(\t )\cr d_ru_s & c_{r,s}(\t
) &{\tilde
b}_{r,s}(\t )\cr}\qquad
\qquad
\bmx{&u_su_r &d_sd_r\cr
u_ru_s&a_{r,s} (\t )&0\cr d_rd_s &0&{\tilde a}_{r,s} (\t )\cr},}
where $\t =\t_1 -\t_2$.

Demanding that the S-matrix \Smat\ commutes with the
supersymmetry generators,
with the action on the states as discussed in \pk , completely
fixes the above
S-matrix elements up to an overall factor:
\eqn\abc{\eqalign{&a=Z_{r,s}(\tht)\sinh ({\tht\over 2}+{i\mu\over
2}(r+s))\cr
&b=Z_{r,s}(\tht)\sinh ({\tht\over 2 }+{i\mu \over 2}(s-r))\cr
&c=Z_{r,s}(\t )ie^{i\mu (r-s)/2}(\sin {r\mu}\sin
{s\mu})^{\half}\cr}
\qquad \eqalign{&{\tilde a}=-Z_{r,s}(\t )\sinh ({\t \over
2}-{i\mu \over
2}(r+s))\cr
&{\tilde
b}=Z_{r,s}(\t )\sinh ({\t \over 2}+{i\mu \over 2}(r-s))\cr
&\tilde c=Z_{r,s}(\tht)ie^{i\mu (s-r)/2}(\sin {r\mu}\sin
{s\mu})^{\half},\cr}}
where $\mu \equiv \pi/n$ and we suppress the $r,s$ dependence of $a,b$ and
$c$. The off-diagonal elements $c$ and $\tilde c$ are convention-dependent,
but $c\tilde c$ is not.  Note that this S-matrix satisfies the ``free
fermion'' condition
\eqn\ff{a(\t ){\tilde a}(\t )+b(\t ){\tilde b}(\t )-c(\t )\tilde
c(\t )=0.}
It also satisfies the Yang-Baxter equation guaranteeing the factorizability:
the elements \abc\ are the R-matrices for the quantum deformation of SU(1,1),
with $q=e^{i\mu}$.

The bootstrap equations, crossing symmetry and unitarity fix the factor
$Z_{r,s}(\tht)$.  In a theory with factorizabilty, S-matrix elements involving
a bound state can be written as the product of S-matrix elements of its
components \rZandZ. The equations describing this property are called the
bootstrap equations.  Two particles $C$ and $D$ have a bound state of mass
$m_{CD}$ ($m_{CD}<m_C + m_D$) when their S-matrix has a pole at $s=m_{CD}^2$,
where
$$s\equiv (p_C+ p_D)^2=m_C^2 + m_D^2 + 2m_C m_D \cosh{\tht},$$
and $\tht=\tht_C-\tht_D$.
In general, there is an $j$-particle bound state of mass $m$
when there is a pole at $(\sum_{i=1}^j p_i)^2 = m^2$.

In our models, a soliton corresponds to a line drawn between two
vertices of a
regular polygon in the $W$-plane. This picture strongly suggests
that these
solitons are bound states of other solitons, because a soliton
$K_{j(j+r+s)}$
can be
continuously deformed into a combination of $K_{j(j+r)}$ and
$K_{(j+r)(j+r+s)}$.  This
motivates our assumption concerning the presence of bound states.
We assume
that $u_2(\tht_1)$ is a bound state of
$u_1(\tht_1 - i\mu)$ and $ u_1(\tht_1 + i\mu)$.  This, of course,
is
consistent with fermion number conservation, since the charge of
$u_r$ is
$r/n$. This means that $a_{1,1}(\tht)$ and $Z_{1,1}(\tht)$ have a
pole at
$\tht=2i\mu$. Notice that the pole from $Z_{1,1}(\tht)$ is
canceled in $\tilde
a_{1,1}$; this is required since there is no particle of charge
$(2/n)-2$ in
the spectrum.

We are now in a position to apply the bootstrap equations to scattering.  The
factorizability requires that scattering a particle with $u_2(\tht_1)$ is the
same as scattering it with each of $u_1(\tht_1+i\mu)$ and $u_1(\tht_1-i\mu)$.
Thus the bootstrap equation for scattering $u_1(\tht_2)$ with $u_2(\tht_1)$ is
$$a_{1,2}(\tht)=a_{1,1}(\tht-i\mu)a_{1,1}(\tht + i\mu).$$
This S-matrix element has an $s$-channel pole at $\tht=3i\mu$, and for $n>3$
this pole corresponds to the bound state $u_3$. The pole at $\tht=i\mu$ is a
$t$-channel pole and does not correspond to another bound state.  Proceeding
in this manner, we find that $u_s(\tht_s)$ for $s<n$ is the bound state of
$u_1(\tht_s-i(s-1)\mu), u_1(\tht_s-i(s-3)\mu)\dots\ u_1(\tht_i+i(s-1)\mu)$.
Thus
\eqn\eboot{a_{1,s}(\tht)=\prod_{j=1}^{s}
a_{1,1}(\tht-i(s+1-2j)\mu).}
In general,
\eqn\ebootii{a_{r,s}(\tht)=\prod_{j=1}^{r} \prod_{k=1}^s
a_{1,1}(\tht-i(r+s+2-2j-2k)\mu).}

By using crossing symmetry, we find an equation for $a_{1,1}$ from \eboot.
Crossing takes, for example, $s\rightarrow r$ and $r\rightarrow n-s$. This
follows from the fact that $u_s$ and $d_{n-s}$ are antiparticles (CP
conjugates).  From the behavior of the above elements under this
transformation we see that
\eqn\cross{Z_{n-s,r}(\t )=Z_{r,s}(i\pi -\t ).}
(Note that the off-diagonal elements $c$ and $\tilde c$ pick up fermion phases
under crossing). Using this in \eboot\ for $s=n-1$ gives
\eqn\eqforY{a_{1,1}(i\pi-\tht)=
{\cosh({\tht\over 2}-i\mu)\over \cosh{\tht\over 2}}
\prod_{j=1}^{n-1}a_{1,1}(\tht +i(2j-n)\mu). }
There are a variety of other constraints one can derive from the
bootstrap
equations, but the supersymmetry makes them all equivalent to
\eqforY.

Another constraint on $Z_{1,1}(\tht)$ comes from the requirement
that the
S-matrix be unitary. This yields
\eqn\unitar{a(\tht)a(-\tht)=1.}
The constraints \unitar\ and \eqforY, along with the requirement
that there
be no additional poles corresponding to bound states, fix
\eqn\forZ{\eqalign{Z_{1,1}(\tht)=&{1\over\sinh ({\tht\over
2}-{i\pi \over
n})}\prod_{j=1}^{\infty}
{\Gamma^2(-{\tht\over 2\pi i}+j) \Gamma({\tht\over 2\pi
i}+j+{1\over n})
\Gamma({\tht\over 2\pi i}+j-{1 \over n})
\over \Gamma^2({\tht\over 2\pi i}+j) \Gamma(-{\tht\over 2\pi
i}+j+{1\over n})
\Gamma(-{\tht\over 2\pi i}+j-{1\over n})}\cr
=&{1\over\sinh ({\tht\over 2}-{i\pi \over n})}\exp(i
\int_{-\infty}^{\infty} {dt\over t} \sin t\t {\sinh ^2{\pi t\over
n}\over
\sinh ^2{\pi t}
}). \cr}}
Using \ebootii\ gives $Z_{r,s}(\t )$ in terms of $Z_{1,1}(\t )$.  We have
thus obtained the exact S-matrix $\Sn$ for \mrp .

We note that if the gamma functions were omitted from $Z_{r,s}$, the elements
$a_{r,s}$ of $\Sn$ \Smat\ would make up the minimal $A_{n-1}$ Toda S-matrix
\rKS, which describes scattering in perturbed ${\bf Z}_{n}$ parafermion
models \rpara.  This is the sense in which our models are the $N$=2 analogs of
the parafermion theories. In fact, these S-matrices were originally introduced
in \rKK\ as supersymmetric generalizations of Toda S-matrices.  However, the
S-matrix $\Sn$ can not be reduced further to a tensor product of a single
basic $N$=2 part with a non-supersymmetric part: the elements \abc\ depend on
the solitons $r$ and $s$.  This is because the $N$=2 algebra depends on the
$\Delta W/m$ structure of the solitons \rWO.  This can also be understood in
terms of the fractional fermion number: the $N$=2 S-matrix elements depend on
the fermion number fractions $r/n$ and $s/n$ of the scattered solitons.  There
is not one basic $N$=2 scattering theory---there is a series $\Sn$.  The $n=2$
case is the sine-Gordon S-matrix at coupling $\beta^2={2\over 3}8\pi$, which
is discussed in detail in \rBL\ and \pk.

\newsec{Thermodynamic Bethe Ansatz}

The thermodynamic Bethe ansatz provides a way of calculating the ground-state
(Casimir) energy $E(R)$ for an integrable quantum field theory on a circle of
length $R$ \refs{\rYY,\tbaref}. It is not an ansatz at all---the only input is
the exact S-matrix.  This provides a useful check on a conjectured S-matrix,
because in the conformal limit, $E(R)$ is proportional to the central charge
of the conformal field theory \rBCN.  In the TBA we find the allowed energy
levels for solitons on a circle of length $L\rightarrow \infty$ at a
temperature $T=1/R$, and fill the levels with solitons so as to minimize the
free energy.  Reversing the roles of space and Euclidean time (we have a
Lorentz-invariant theory) we can extract the Casimir energy $E(R)$ in terms of
integral equations.  The results are of the general form
\eqn\fe{E(R)=-\sum _a {m_a\over 2\pi }\int d\t \cosh \t \ln
(1+e^{-\e _a(\t
)}),}
where the $\e _a(\t)$ are solutions to the coupled integral
equations:
\eqn\TBA{\e _a(\t )=m_aR\cosh (\t )-\sum _b \int {d\t ^{\p}\over
2\pi}\phi
_{ab}(\t -\t ^{\p})\ln(1+e^{-\e _b(\t ^{\p})}).}
When the S-matrix is diagonal, the $m_a$ are the soliton masses and
$\phi_{ab}= i{d\over d\tht} \ln S_{ab}(\tht)$. Obtaining the correct $m_a$ and
$\phi _{ab}$ for a theory with a non-diagonal S-matrix such as ours is more
involved: it requires ``diagonalizing'' a many-soliton state so that the
action of bringing one soliton through all the others is diagonal
\refs{\rAlZii,\pk}.  We define a transfer matrix with
 Boltzmann weights given by the S-matrix elements.  In order to obtain the
$m_a$ and the $\phi _{ab}$ in \TBA , we need to know the eigenvalues of the
transfer matrix with periodic boundary conditions.  Our S-matrices, by fermion
number conservation, are of the 6-vertex type, allowing us to use the
algebraic Bethe ansatz to obtain the transfer-matrix eigenvalues and, thereby,
the $\phi _{ab}$ in \TBA .

The result we derive in this section is that each $N$=2 TBA system is a
generalization of a TBA system for minimal affine Toda S-matrices \rKMi.  Our
TBA system has $n-1$ massive particles like the Toda system, with the same
masses and the same $\phi_{ab}$. Our system also has two massless
pseudoparticles which couple to the Toda particles. These massless particles
account for the entropy of the transfer-matrix eigenvalues.

To do the thermodynamics, we have a large number $N$ of particles distributed
along a circle of length $L$. Since the soliton type labels $r$ and $s$
scatter diagonally, the eigenvectors of the transfer matrices have definite
numbers $N_r$ of particles of type $r$ (i.e., the number of $u_r$ particles
plus the number of $d_r$ remains conserved). We thus define a distribution
$P_{r}(\t )$ of allowed rapidities for species $r$ (whether it be $u_r$ or
$d_r$), for $r=1,\dots ,n-1$.  We, likewise, define $\rho _{r}$ to be the
distributions of rapidities actually occupied by solitons of type $r$.  The
distributions $P_{r}(\t )$ are related to the $\rho_r (\tht)$ by the
requirement that the wavefunction be single-valued upon bringing such a
soliton around the circle:
\eqn\Pdist{2\pi P_r(\t )=m_rL\cosh \t +Im {d\over d\t}\ln\l _r(\t
),}
where $\l _r (\t )$ is the distribution of eigenvalues of the transfer matrix
for bringing a solitons of type $r$ through the others.

For a given $(r,s)$, our S-matrices \Smat\ are of the 6-vertex type (with $u$
and $d$ corresponding to up and down (or left and right) arrows,
respectively).  We can thus use the algebraic Bethe ansatz \rFST\ to find the
$2^N$ eigenvalues of the transfer matrices. Generalizing the discussion in our
last paper \pk\ to our present situation, we represent the eigenvectors of the
transfer matrix for bringing a soliton of type $r$ (i.e. $u_r$ or $d_r$)
through the other solitons on the circle as:
\eqn\evects{\psi _r=\prod _{k=1}^m B_p(y_{pk})|\prod
_{s=1}^{n-1}\prod
_{j_s=1}^{N_s}d_s(\t _{j_s})\>,}
where $m$ is any number less than $N=\sum N_s$, $B_p(y_{pk})$ is the transfer
matrix for bringing a soliton of type $u_p(y_{pk})$ through the other solitons
and ending up with a soliton of type $d_p(y_{pk})$, and the allowed values of
$y_{pk}$ are determined as in \pk.  Again, the scattering is diagonal in the
$r,s$ labels: when bringing a solitons of type $p$ through the others, we end
up with one of type $p$. We will see that any particular $p$ can be used in
\evects .  The discussion in the appendix of \pk\ applies to the scattering of
a particular $r$ and $s$; the generalization to our present case of scattering
with several types of $r$ and $s$ is straightforward.  We obtain the transfer
matrix eigenvalues in terms of the $y_k$.  In particular, the desired
ingredients in \Pdist\ are expressed as:
\eqn\Imdlog{
\eqalign{&Im {d\over d\t}\ln \l _{r}(\t )\cr
&=\sum _{s=1}^{n-1}\int d\t '\rho _{s}(\t ')Im
{d\over d\t }\ln Z_{r,s}(\t -\t ')+Im{d\over d\t}\sum _{k=1}^m\ln
\sinh ({\t-
z_k+ir\mu \over 2})
\cr
&+Im{d\over d\t}\sum _{k=m+1}^{N}\ln \sinh ({\t -z_k-ir\mu\over
2}),\cr}}
where $Z_{r,s}(\t )$ is given by \forZ\ and the bootstrap \ebootii .  The
$z_k$ in \Imdlog\ are the solutions of
\eqn\zs{\prod _{s=1}^{n-1}\prod _{j_s=1}^{N_s}{\sinh ((z-\t
_{j_s}
+is\mu)/2 )\over -\sinh ((z-\t _{j_s}-is\mu)/2)}=(-1)^{N+1}}
(the $y_{pk}$ in \evects\ are given by $z_k+i\mu p$). This is a polynomial
equation of order $N$ in the $e^{z_k}$, so there are $N$ solutions.  As in
\pk\ each $z_k$ appears exactly once in the $N$ terms of \Imdlog.  The $2^N$
different eigenvalues correspond to the option of putting a given solution
$z_k$ in the first or second term in \Imdlog. Notice that the $z_k$ and hence
the eigenvalues are independent of the choice of $p$.

The solutions of \zs\ are of the form $z=z_0$ and $z=z_{\bar 0}+i\pi$, with
$z_0$ and $z_{\bar 0}$ real.\foot{It is worth noting that the free fermion
condition \ff\ makes our analysis easier than that of the general six-vertex
model: there are no string-type solutions to \zs.} Defining corresponding
distributions $P_l(z_l)$, $l=0, \bar 0$, taking the log of \zs\ yields
\eqn\Pls{2\pi P_l(z_l)=\sum
_{s=1}^{n-1}\sin {s\mu}\int d\t' {\rho _s(\t' )\over \cosh
(z_l-\t' )-a_l\cos
{s\mu}},}
where $a_0=-a _{\bar 0}=1$.  Taking the integral of $P_0 + P_{\bar 0}$ with
respect to $z$ shows that we have $N$ different values of z in the
thermodynamic limit:
$$\int dz_0 P_0(z_0) +\int dz_{\bar 0} P_{\bar 0} (z_{\bar
0})=\sum_{s=1}^{n-1} \int d\tht' \rho_s(\tht')= N.$$
This confirms that we have all the $z_k$ needed for the $2^N$
eigenvalues in \Imdlog.

We define $P^{+}_0(z_0)$ and $P^{-}_0(z_0)$ to be the distributions
corresponding to the $k> m$ and $k\leq m$ terms in
\Imdlog , and vice-versa for $l=\bar 0$. Thus $P_l=P^+_l +P^-_l$.
In terms of these distributions, \Imdlog\ becomes
\eqn\ImdlogII{
\eqalign{&Im {d\over d\t}\ln \l _{r}(\t )
=\sum _{s=1}^{n-1}\int d\t '\rho _{s}(\t ')Im
{d\over d\t }\ln Z_{r,s}(\t -\t ')\cr
&+\sum _{l=0,\bar 0}\int {dz_l\over 2}\sin
{r\mu}{P^+_l(z_l)-P^-_l(z_l)\over \cosh
(z_l-\t )-a_l\cos {r\mu}}.\cr}}
Writing $P_l^- =P_l-P_l^+$
in \ImdlogII\ and using \Pls , we have for \Pdist
\eqn\Pyuck{
\eqalign{&2\pi P_r(\t )-m_rL\cosh \t=\cr
&\sum _{l=0,\bar 0}\int dz_l \phi_{r,l} P^+_l(z_l)+
\sum _{s=1}^{n-1}\int d\t '\rho _s(\t ')\phi_{r,s}(\t -\t')
,\cr}}
where
%
\eqn\phiab{\phi _{r,l}(\t )={\sin (r\mu)\over
\cosh (\t )- a_l\cos (r\mu )},}
\eqn\phiabii{\phi_{r,s}(\t)=Im {d\over d\t}\ln Z_{r,s}(\t)-F_{
r,s}(\t)-F_{n-r,n-s}(\t),}
and
\eqn\Fis{2F_{r,s}(\t)=\int
dt\cos (t\t ){\sinh rt\mu \sinh st\mu \over \sinh ^2(\pi t)},}
as can be seen from a fourier transform.  It follows from a bit of algebra
that the $\phi_{r,s}$ in \phiabii\ can be expressed as
\eqn\forphi{\phi_{r,s}(\t )=Im {d\over d\t}\ln A_{r,s}(\t),}
where the $A_{r,s}$ are the $a_{r,s}(\t )$ of \Smat\ with the gamma
functions from \forZ\ removed: i.e.,
\eqn\ars{A_{r,s}[n](\t)=\prod _{j=1}^r\prod _{k=1}^s{\sinh
({\t\over
2}+i\mu(j+k-
{r+s\over 2}))\over \sinh ({\t\over 2}+i\mu(j+k-2-{r+s\over
2}))}.}
The $\phi _{r,s}$ given by \forphi\ are precisely the $\phi_{r,s}$ for the
minimal $A_n$ Toda S-matrix \rKMi\ which can be conveniently written as \rKar
\eqn\forphii{\phi_{r,s}(\t)=\int {dt\over 2\pi} e^{it\t}
\left(\delta_{rs}-2{\cosh \mu t \sinh(\pi-r\mu)t \sinh s\mu t \over \sinh \pi
t\sinh \mu t}\right)}
for $r\geq s$ with $\phi _{sr}=\phi _{rs}$.

We minimize the free energy with respect to the $\rho _s$ and $P_l^+$, subject
to the constraints \Pls\ and \Pyuck .  Defining
\eqn\pseudos{{\rho _s(\t)\over P _s(\t )}={e^{-\e _s(\t)}\over
1+e^{-\e _s(\t
)}}\quad {P^+_l(y_l)\over P_l(y_l)}={e^{-\e _l (y_l)}\over
1+e^{-\e _l
(y_l)}},}
for $s=1,\dots n-1$ and $l=0,\bar 0$, we obtain a TBA system of coupled
integral equations of the type \TBA\ and \fe .  The species $a$ and $b$ in
\TBA\ run over the labels $s=1,\dots ,n-1$ and $l=0,\bar 0$.  Species $s$ has
$m_s=M\sin s\mu$ while species $l$ are massless: $m_l=0$.  The elements $\phi
_{ab}(\t )$ in \TBA\ are given by \forphii\ and \phiab .  Most other TBA
systems have $\phi _{ab}$ which can be simply encoded by a Dynkin diagram of a
(perhaps extended) simply-laced Lie algebra; examples of such systems will be
seen in sects.6 and 7.  This TBA system does not immediately have such an
interpretation.

To escape from all these equations, we here describe the $W=X^4/4- X$ ($n=3$)
case explicitly. (The $n=2$ case was discussed in detail in \pk.)  The vacua
form a triangle in $X$ space. A doublet ($u_1,d_1$) of solitons interpolate
between vacua going counterclockwise. These particles have fractional charges
of ($1/3, -2/3$) respectively.  Their antiparticles, ($u_2,d_2$), interpolate
clockwise and have fractional charges of ($2/3,-1/3$). All particles here have
the same mass, since the distance $|\Delta W|$ is the same for all solitons.
In the associated TBA system, there is one massive particle for every doublet
($u_r,d_r$). Thus the TBA system for $n=3$ consists of two massive particles
$a=1,2$ with $m_1=m_2=M$ and two massless particles $a=0,\bar 0$.  The
elements $\phi_{ab}(\t)$ are given by \phiab\ and \forphi, and
$\phi_{1,1}=\phi_{2,2}$ and $\phi_{1,2}$ are those which arise in the
three-state Potts model (${\bf Z}_3$ parafermions) \tbaref.

\subsec{The UV and IR limits}

In the UV limit ($MR\rightarrow 0$) the perturbing parameter in \mrp\ goes to
zero and we return to the conformal theory.  The Casimir energy $E(R)$ \fe\ in
this limit must then equal that of the conformal theory on a periodic box of
size $R$: $E(R)=-\pi c/6R$, where $c$ is the central charge \rBCN. (For
nonunitary theories $c$ is replaced by $c_{eff}=c-12(h_{min}+\bar h_{min})$).
Extracting this result from our scattering theory via
the TBA gives a very powerful check on the spectrum and S-matrix.  In
particular, it rules out the inclusion of additional particles---the one
possibility not excluded by the exact S-matrix constraints of section 3 (i.e.\
the CDD ambiguity).

The Casimir energy can be explicitly calculated in the UV limit
\refs{\rKR,\tbaref}:
\eqn\uvlim{E(m_aR\rightarrow 0)\sim -{1\over \pi R} \sum _a
[{\cal
L}({x_a\over
1+x_a})-{\cal L}({y_a\over 1+y_a})],}
where ${\cal L}(x)$ is Rogers dilogarithm function\rLewin
$${\cal L}(x)=-{1\over 2}\int_0^x dy\left[{\ln y\over (1-y)} +
{\ln(1-y)\over
y}\right],$$
$x_a=\exp(-\epsilon_a(0))$, and $y_a=\exp(-\epsilon_a(\infty))$. It
follows from
\TBA\ that the constants $x_a$ are the solutions to the equations
\eqn\xa{x_a=\prod _b (1+ x_b)^{N_{ab}},}
where $N_{ab}={1\over 2\pi}\int d\t \phi _{ab}(\t )$.  The constants $y_a$ in
\uvlim\ are nonzero only for those species $a ^{\p}$ with $m_{a^{\p}}=0$,
where they are the solutions to
\eqn\ya{y_{a^{\p}}=\prod _{b^{\p}} (1+
y_{b^{\p}})^{N_{a^{\p}b^{\p}}},}
where $b^{\p}$ also runs only over massless species.

In our case, the $N_{ab}$ obtained from \phiab\ and \forphii\ are given by
\eqn\Nab{
\eqalign{N_{ll^{\p}}=0\qquad
N_{r0}&=1-{r\over n}\qquad N_{r\bar 0}={r\over n}\cr
N_{rs}&= -(2s(1-{r\over n})-\delta_{rs})\qquad (\hbox{for } r\ge
s),\cr}}
and the others follow from $N_{sr}=N_{rs}$.  The not-at-all-obvious solution
to \xa\ is
\eqn\xamostrel{\eqalign{x_0=x_{\bar 0}=&{\sin{3\pi\over
2(n+1)}\over\sin{\pi\over 2(n+1)}}\cr
x_r=&{\sin^2{\pi\over (n+1)}\over \sin{(2r+3)\pi\over 2(n+1)}
\sin{(2r-1)\pi\over 2(n+1)}},\cr}}
and to \ya\ is
\eqn\yamostrel{y_0=y_{\bar 0}=1;\qquad y_r=0.}
We note that this is a new type of TBA system.

Amazingly, there are dilogarithm identities which can be used to evaluate
exactly the expression \uvlim , with the above $x_a$ and $y_a$, for the UV
limit of the free energy. From the appendix of \rKR\ (correcting a typo in
equation (A2.4)) and equation (5.139) of \rLewin, we have derived that
\eqn\dilogid{2{\cal L}\left(1-{\sin^2{\pi\over 2(n+1)}\over
\sin^2{\pi\over
(n+1)}}\right) + \sum_{r=1}^{n-1}{\cal L} \left({\sin^2{\pi\over
(n+1)}\over
\sin^2{(2r+1)\pi\over 2(n+1)}}\right)
= {\pi^2\over 6}(4-{6\over n+1}).}
Since ${\cal L}(1/2)=\pi^2/12$, the contribution from $y_0$ and
$y_{\bar 0}$ to \uvlim\ subtracts $\pi^2/6$ from this. The final result is
\eqn\rocknroll{E(R)=-{\pi\over 6R}\left({3 (n-1) \over
(n+1)}\right),}
in agreement with the appropriate central charge!

The leading term for the free energy in the infrared limit
($mR\rightarrow\infty$) is easily found from our TBA system to be
\eqn\IR{E(mR\rightarrow \infty)=-\sum _{r=1}^{n-1}\int {d\t\over
2\pi}2m_r\cosh \t e^{-m_aR\cosh \t} .}
In this limit the system becomes a dilute gas of free particles and, thus,
\IR\ is the expected answer; the factor of two comes from our soliton doublets
$(u_r,d_r)$. In the TBA, this factor arises from the pseudoparticles.

\newsec{The General Structure}

In this section we extend our analysis of perturbed integrable $N$=2 theories,
in direct analogy with an extended class of integrable perturbed $N$=0 models.
These $N$=0 models at their critical points are described by the cosets
\eqn\coset{{SU(n)_k \otimes SU(n)_1 \over SU(n)_{k+1}}.}
With a particular perturbation, these models remain integrable (see \rABL\ for
details). For $n$=2, the coset models are the minimal models, and the
perturbation is by the $\Phi_{1,3}$ operator \rZamint. As is well-established,
the spectrum in this case consists of equal-mass solitons interpolating
between adjacent wells of a potential with $k+1$ degenerate wells \RSOS. For
$k$=1, the coset models are ${\bf Z}_n$ parafermions. After perturbation, the
spectrum is that of $SU(n)$ affine Toda theory: there are $n-1$ particles with
mass proportional to $\sin (r\pi/n)$, where $r=1\dots n-1$. The scattering is
diagonal and given by the minimal Toda S-matrix \rpara. For general $n$ and
$k$, the spectrum consists of solitons interpolating between the wells of a
more complicated potential \rABL. The masses in the spectrum depend only on
$N$, not on $k$. In this section we will display an appropriate potential for
these solitons: the bosonic part of the analogous $N$=2 Landau Ginzburg
superpotentials to be discussed.

The $N$=2 analog of theory \coset\ has the Kazama-Suzuki \rKzS\
coset
description \lwv
\eqn\susycoset{{SU(n)_k \otimes SO(2(n-1))_1\over
SU(n-1)_{k+1}\otimes U(1)},}
where the $SO(2(n-1))_1$ is the contribution of the fermions added for
supersymmetry. This conformal theory is believed to remain integrable when
perturbed in a particular flat direction (in the sense of \dvv ) by one of its
relevant chiral primary operators \rken.  One reason for this belief is that
the differential equations obtained in \rCV\ for the renormalization group
flow of the Ramond ground state metric along this perturbation are classical
$SU(n)$ affine Toda differential equations---a major simplification of the
equations that ordinarily characterize the metric.  We will see that the
S-matrices for arbitrary $n$ and $k$ have an $N$=2 part (the S-matrix of
section 3) tensored with a part arising from the soliton structure, and that
the soliton structure is the same as that of the associated $N$=0 integrable
theory obtained by perturbing \coset .

The effective Landau-Ginzburg superpotential characterizing our integrable
perturbation of theory \susycoset\ is given by the $k$-th generalized
Chebyshev polynomial in $n-1$ variables.  Define
\eqn\xt{X(t)=\sum _{i=0}^{n}X_it^i=\prod _{i=1}^{n}(1+q_it),}
with $X_0=1$ and $X_{n}=\beta$.  The $k$-th generalized Chebyshev
superpotential in $n-1$ variables is given by the coefficient of
$t^{n+k}$ in
\eqn\wt{W(t)=-\ln(X(-t))}
written in terms of the variables $X_i$ \rGep.  In other words,
$$W(X_i)=\sum _{i=1}^{n}{q_i^{n+k}\over n+k},$$
subject to $\prod _{i=1}^{n}q_i=\beta$, and written in terms of the $X_i$
defined in \xt . These models were shown in \rGep\ to have the $SU(n)_k$
fusion rules as the chiral ring structure constants.  These theories have, for
any $k$, a \Zn\ symmetry corresponding to $X_i\rightarrow \omega ^iX_i$ with
$\omega ^n=1$.  This symmetry will be reflected in the soliton masses, fermion
number fractions, and the $N$=2 part of the S-matrix: the soliton structure is
a tensor product of that of the basic \Zn\ $N$=2 theory \mrp\ with an
additional $N$=0 structure.

For $n$=2 these are the Chebyshev theories discussed in \pk. The potential has
$k+1$ degenerate wells, just as in the $N$=0 models \coset\ with $n$=2.  The
solitons interpolate only between adjacent wells---any other potential soliton
decomposes into these.  The S-matrices for \susycoset\ in these cases are
equal to the S-matrices for \coset\ tensored with the $N$=2 part
$S^{N=2}[n=2]$ \refs{\rBL,\pk}.  The $k=1$ Chebyshev theory in $n-1$
variables \wt\ is easily seen to be equivalent to the basic \Zn\ integrable
theory \mrp\ of sec.\ 2.2, where $X=X_1$ and $X^r=X_r$.  While we saw in
sect.\ 3 that the S-matrix is not quite a tensor product of the analog $N$=0
S-matrix for \coset\ with an $N$=2 part, the two theories are closely related:
they have the same mass spectrum and the same bound-state structure.

To find the soliton structure for the general models, we find the minima of
the bosonic potential $\sum_i |\del W/\del X_i|^2$. These minima are the
$(n+k-1)!/(n-1)!k!$ critical points of the above superpotential, and can be
labelled by the highest weights ${\bf \mu}=\sum _{i=1}^{n-1}n_i\Lambda _i$ of
$SU(n)_k$, where the $\Lambda_i$ are the fundamental weights \rGep. In other
words, the $n_i$ are nonnegative integers such that $\sum _i n_i\leq k$.  It
can be shown that the value of the superpotential at the critical point with
label $\bf
\mu$ is
\eqn\watmu{W(X_i^{({\bf \mu})}) =\omega ^{b({\bf \mu})}{n\over
n+k},}
where $\omega ^{n}=1$ and $b({\bf \mu })$ is given by $\sum
_{i=1}^{n-1}il_i$
(i.e.\ the number of boxes in the Young tableaux for the
$SU(n)_k$ highest
weight $\bf \mu$).

A soliton $K_{\mu\nu}$ interpolates between the vacua $\mu$ and $\nu$. In
analogy with RSOS statistical-mechanical lattice models \rJKMO, we assume that
only a subset of the solitons are fundamental---the others decay into these.
In this analogy, the vacua, each labelled by a representation of $SU(n)_k$,
are the ``heights'' of the lattice model. Each soliton $K_{\mu\nu}$ is a
``link'' variable of the lattice model.  In the lattice model these links must
be the fundamental weights $\Lambda_r$ of the affine algebra
$\widehat{SU(n)}$.  This leads us to our assumed restriction: a fundamental
soliton exists between two vacua $\mu$ and $\nu$ iff the representation
associated with $\nu$ is in the product of a fundamental representation
$\Lambda_r$ with the representation $\mu$.  This assumption is the obvious
extension of the results for $n=2$ and for $k=1$ and can probably be
proven along the lines of \rLW\ by using the W-algebra structure.

So supposing that our $SU(n)_k$ representations $\mu$ and $\nu$
satisfy $\mu
\otimes \Lambda _r=\nu +\dots $, where $r=b(\nu )-b(\mu )$ mod
$n$ $(1\leq
r\leq n-1)$, there is a soliton $K_{\mu\nu ;r}$ of type
$r=1,\dots ,n-1$
connecting vacuum $\mu$ with vacuum $\nu$.  As before, soliton
$K_{\mu\nu ;r}$
is a $(u_r,d_r)$ doublet with the Toda masses $m_{r}$=$|\Delta
W|$=$m\sin
(\pi r/n)$ and with fermion number
$(f_r,f_r-1)$ where
\eqn\fract{\eqalign{f_{r}&=-{1\over 2 \pi}\Delta Im \ln \det
({\partial ^2 W\over
\partial X_i\partial X_j})\cr
&={r \over n}.\cr}}
The second equality can be shown via relations from \refs{\rken
,\bonus}.

To give an example, the lattice model for the $n$=2 case is the simplest RSOS
model, with $k+1$ heights numbered $1\dots k+1$, and the restriction that
heights on adjacent lattice sites differ by $\pm 1$. In the soliton language,
this means that fundamental solitons connect only adjacent vacua.  The two
fundamental weights of $\widehat{SU(2)}$ correspond to a soliton either going
up or going down one vacuum. Another simple example is the $k$=1 case, where
there is no restriction---any vacuum can be reached from any other. In this
case, there are $n-1$ fundamental weights, so there are $n-1$ solitons. This
is the structure of the models of sections 2--4, or of the affine $SU(n)$ Toda
S-matrix.

If we forget the supermultiplets, this general structure is exactly the same
as that conjectured for the perturbed coset models \coset \rABL.  Conservation
of the higher-spin currents requires that the scattering is diagonal in the
soliton type labels $r$ and $s$, just like the $k=1$ case of section 3. Thus
the general scattering process is of the form
\eqn\scatt{
K_{\mu\nu;r}(\t _1)K_{\nu\sigma;s}(\t _2)\rightarrow
S(\mu\nu\sigma\lambda;n,k)(\t )K_{\mu \lambda;s}(\t
_2)K_{\lambda \sigma;r}(\t _1).}
Factorizability requires that the S-matrix elements in \scatt\ are related to
the lattice-model Boltzmann weights $w(\mu\nu\sigma\lambda ;n,k)$ given in
\rJKMO, up to an overall $\tht$-dependent constant.  Since the scattering is
diagonal in the labels $r$ and $s$ we can multiply the S-matrix by any
function of these labels and the result will still satisfy the Yang-Baxter
equation. In \rABL\ this function was referred to as the `CDD' part of the
S-matrix; it is natural to take this function to be the $A_{r,s}[n]$ of \ars,
which is the S-matrix of minimal affine $SU(n)$ Toda theory. Thus the S-matrix
element for the process \scatt\ in the theory
\coset\ is
\eqn\sabl{S^{N=0}(\mu\nu\sigma\lambda;n,k)(\t)=
w(\mu\nu\sigma\lambda ; n,k)(\t) A_{r,s}[n](\t ),}
with the overall factor in $w$ to be fixed by unitarity.

The particle structure of the $N$=2 analog \susycoset\ is more complicated
because every soliton $K_{\mu\nu;r}$ forms a $(u_r,d_r)$ supermultiplet.
However, it is easy to generalize \sabl\ by analogy with the $k=1$ case. Here
the Boltzmann weights $w(n,1)$ are all $1$ and the S-matrix in the $N$=0 case
is diagonal and equal to $A_{r,s}[n]$. We showed in section 3 and confirmed in
section 4 that in the $N$=2 case with $k$=1 the S-matrix is
$S^{N=2}_{r,s}[n]$, as defined in \Smat, \abc\ and \forZ\ . Thus our
conjecture for the S-matrix for process \scatt\ in the general $N$=2 case is
\eqn\ssusy{S^{N=2}(\mu\nu\sigma\lambda ;n,k)(\t
)=w(\mu\nu\lambda\sigma
;n,k)(\t )\otimes S^{N=2}_{r,s}
[n](\t),}
of the general form \sprod . This S-matrix is consistent with the
dependence
of the superalgebra on $\Delta W_{\mu\nu}$, which is found using
\watmu.

We have little doubt that these are the correct S-matrices, although we have
not confirmed them by using the thermodynamic Bethe ansatz to calculate the
ground-state energy. Such an analysis was done for the $k$=1 case in sect.\ 4
and for the $n$=2 case in \pk. The missing ingredient in the general case
is the diagonalization of the transfer matrices for the $N$=0 part of
\ssusy ; if one were to obtain
the TBA system for
\sabl , it would be straightforward to obtain the TBA system for \ssusy .
In the $n$=2 case,
the TBA system of equations for the $N$=0 (RSOS$)_k$ part of \ssusy\
was conjectured in
\RSOS , with the cases $k=1,2$ proven there and with the rest proven in \un.
Generalizing this to arbitrary $n$ would require an analysis along the lines
of \rBR\ to obtain the transfer matrix eigenvalues for the $N$=0 part of
\ssusy.
(Actually, another generalization using \rKuR\ is required, because
\rBR\ covers only scattering of particles with a fixed $r$.)

\newsec{Some Sigma Model Thermodynamics}

Supersymmetric sigma models on $CP^{n-1}$ are very closely related to the
perturbed LG theory \mrp .  Since $CP^{n-1}$ is Kahler, the supersymmetric
sigma model has $N$=2 supersymmetry but since $CP^{n-1}$ is not Ricci flat
(the first Chern class is not zero), the sigma model is not conformally
invariant.  The supersymmetric $CP^{n-1}$ sigma model is, however, integrable.
While integrability of the non-supersymmetric $CP^{n-1}$ sigma models is
spoiled for $n\neq 2$ by anomalies, such anomalies cancel in the
supersymmetric case \intcpnrefs .  To see a connection with the perturbed LG
theory \mrp\ note that if we denote the Kahler form of $CP^{n-1}$ by $X$, then
the classical cohomology ring of $CP^{n-1}$ is generated by powers of $X$ with
the ideal $X^n=0$, where powers are wedge products.  Instantons modify this to
the quantum deformed cohomology ring generated by powers of $X$ with the
modified ideal $X^n=\beta =e^{-S_{INST}}$ where $S_{INST}=R+i\tht$ is the
holomorphic instanton action \refs{\rNar,\rken}.  This is precisely the ring
of the deformed theory \mrp .  (The twisted, topological versions \witten\ of
the two theories are identical \refs{\qrings , \rken}.)  Although it appears
that there are no solitons in supersymmetric $CP^{n-1}$, the fundamental
particles are in fact solitons resulting from a spontaneously broken ${\bf
Z}_n$ symmetry \witN, just as solitons in \mrp\ interpolate between vacua
corresponding to a spontaneously-broken ${\bf Z}_n$ symmetry.

The two theories, of course, {\it are} different; for example, the UV (flat)
limit of the $CP^{n-1}$ sigma model has central charge $3(n-1)$ whereas the UV
limit of \mrp\ is the minimal model with central charge $3(n-1)/(n+1)$.  In
the IR limit both theories are massive, flowing to $c=0$.  In \rCVii\ it was
found that the differential equations for the metric on the space of Ramond
ground states as a function of $\beta$ are the same for the two theories
(namely the classical affine $\widehat A _{n-1}$ Toda equations), the only
difference being in the boundary condition at $\beta\rightarrow 0$.

The soliton content and S-matrix of supersymmetric $CP^{n- 1}$ have been
studied in a variety of works \refs{\witN ,\cpnrefs}. The soliton content of
the $CP^{n-1}=SU(n)/SU(n-1)\otimes U(1)$ model exhibits the $SU(n)$ structure:
for $r=1,\dots n-1$ there are $n!/r!(n-r)!$ solitons corresponding to the
fundamental weight representation $\Lambda _r$ of $SU(n)$.  Each of these
solitons is a doublet $(u_{r\alpha},d_{r\alpha})$ under the $N$=2
supersymmetry with $r=1,\dots n-1$, and $\alpha =1,\dots n!/r!(n-r)!$.  In
\cpnrefs\ the solitons were termed bosons and fermions, but the standard
arguments show that the fermion numbers of the $N$=2 multiplets are those of
\mrp , namely multiples of $1/n$.

After some algebra, the results of \cpnrefs\ for the scattering of the $r=1$
solitons can be rewritten as a tensor product \sprod\ of the basic $N$=2
supersymmetric part corresponding to theory \mrp\ with an additional $N$=0
structure:
\eqn\CPSmat{\bmx{&d_{1\gamma}u_{1\delta}&u_{1\gamma}d_{1\delta}\cr
u_{1\alpha}d_{1\beta}&b_{1,1}(\t )&{\tilde c}_{1,1}(\t )\cr
d_{1\alpha}u_{1\beta} & c_{1,1}(\t ) &{\tilde
b}_{1,1}(\t )\cr}P_{\alpha\beta}^{\gamma\delta}(\t)\qquad
\bmx{&u_{1\gamma}u_{r1\delta} &d_{1\gamma}d_{1\delta}\cr
u_{1\alpha}u_{1\beta}&a_{1,1} (\t )&0\cr d_{1\alpha}d_{1\beta}
&0&{\tilde a}_{1,1} (\t )\cr}P_{\alpha\beta}^{\gamma\delta}(\t),}
where $\t =\t_1-\t_2$, the elements
$a_{1,1},b_{1,1},c_{1,1},\dots $ are
precisely those given
by \abc , and
\eqn\Pis{P_{\alpha\beta}^{\gamma\delta}=
{\tht\over \tht-{2\pi i\over n}}{\Gamma(-{\tht\over 2\pi i})
\Gamma({\tht\over 2\pi i}+{1\over n})
\over \Gamma({\tht\over 2\pi i}) \Gamma(-{\tht\over 2\pi
i}+{1\over n})}
(\delta _{\alpha\delta}\delta _{\beta\gamma}-{2\pi i \over
n\theta}\delta_{\alpha\gamma}\delta _{\beta\delta}).}
We see that for scattering of the basic solitons ($r=1$), the $CP^{n-1}$
scattering theory is the tensor product of the perturbed minimal model \mrp\
scattering theory with the additional $\alpha$ soliton labels which scatter as
\Pis .  The scattering of solitons with general $r,s$ will be of
the form of \abc\ with the prefactor and scattering of additional soliton
labels governed by the bootstrap and $SU(n)$ structure.  Thus the
supersymmetric $CP^{n-1}$ sigma model S-matrix is of the form \sprod .

In fact, the soliton structure and S-matrices of supersymmetric $CP^{n-1}$ are
precisely given by the $k\rightarrow \infty$ limit of the integral
perturbation of \susycoset\ discussed in the last section; i.e. the
$k\rightarrow \infty$ limit of the $(n-1)$-variable Chebyshev theory.  In that
limit the quantum-group restriction on the allowed kinks for the $N$=0 part
goes away and we obtain the full $SU(n)$ structure of the $N$=0 part of
supersymmetric $CP^{n-1}$.  The $N$=2 structure of the theories, again, is
always the same, being that of the minimal theories
\mrp .  For example, in the $n$=2 case, vacua are labelled by all the integers
(no maximum or minimum) just as in the sine-Gordon model, and the soliton
doublets connect adjacent vacua. One can view this as a theory with only four
solitons---the doublets $K_{j,j-1}$ and $K_{j,j+1}$.  These correspond to the
particles in \CPSmat, and to the fields of the supersymmetric $CP^1$
Lagrangian (i.e.\ the coordinates of the supermanifold).  This similarity to
sine-Gordon is no coincidence---these models are equivalent to the $N$=2
super-sine-Gordon theories and their affine Toda generalizations at a critical
value of the coupling, as we will discuss in sect.\ 7.

Supersymmetric sigma models on the Grassmannian manifolds
\eqn\grass{Gr(p,n-p)={U(n)\over U(p)\otimes U(n-p)}}
are also integrable, $N$=2 supersymmetric theories; the case $p$=1 is the
$CP^{n-1}$ sigma model.  Integrability and relations for the exact S-matrices
of these theories were discussed in \grrefs .  For any $p$, the solitons in
theory \grass\ correspond to a spontaneously broken \Zn\ symmetry.  This \Zn\
structure is exhibited in the instanton--modified cohomology ring of theory
\grass\ conjectured in \refs{\qrings ,\rken}; the \Zn\ is associated with the
first Chern class of the manifold \grass .  We expect the soliton structure in
theory \grass\ to be the tensor product of that of theory \mrp\ with an
additional $p$ dependent $N$=0 structure, with an S-matrix of the form
\sprod .  This expectation is consistent with the relations obtained in
\grrefs\ for the exact S-matrix of theory \grass .

\subsec{TBA for $CP^1$}

In order to obtain the TBA system of equations for the $CP^{n-1}$
supersymmetric sigma model we will have to find the eigenvalues of the
transfer matrix for the scattering \Pis\ of the $\alpha$ labels; the remaining
contribution to the full transfer matrix eigenvalues having been obtained in
sect.\ 4 in the context of theory \mrp .  While these eigenvalues can be
obtained by a generalization of the algebraic Bethe ansatz \rKuR, we will
limit our consideration to the case of $CP^1$ where the labels in
\Pis\ run
over two values and, thus, the ordinary algebraic Bethe ansatz suffices to
find the transfer matrix eigenvalues.  A very similar TBA analysis arose
recently in another context \newZ , where a scattering theory of massless
particles of the same chirality (with trivial left-right scattering) was
conjectured to describe the conformal field theory $SU(2)_1$.

By the above discussion $CP^1$ corresponds to the $k\rightarrow\infty$ limit
of the Chebyshev theory in one variable; i.e. the $k
\rightarrow\infty$ limit
of the theories discussed in \pk .  The algebraic Bethe ansatz for the $n=2$
case of \Pis\ \un , along with our analysis of the $n=2$ case of
\Smat\ \pk ,
confirms that the Casimir energy of supersymmetric $CP^1$ is described by the
diagram
\bigskip
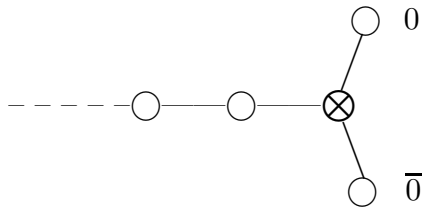
\centerline{\hbox{\rlap{\raise27pt\hbox{$\hskip4.25cm\bigcirc\hskip.25cm
0$}}
\rlap{\lower27pt\hbox{$\hskip4.1cm\bigcirc\hskip.3cm \bar 0$}}
\rlap{\raise14pt\hbox{$\hskip3.9cm\Big/$}}
\rlap{\lower14pt\hbox{$\hskip3.8cm\Big\backslash$}}
-- -- -- -- --$\bigcirc$------$\bigcirc$------{$\bigotimes$} }}
\bigskip

\noindent
where each node labels a particle in the TBA system \TBA.  The open nodes are
zero-mass pseudoparticles (there are an infinite number here), $\otimes$
labels the massive particle, and $\phi_{ab}=1/\cosh\tht$ for nodes connected
on the diagram, and zero otherwise.  The UV limit, of course, gives the
correct $c=3$ as it is the $k\rightarrow\infty$ limit of the computation in
\pk, where we obtained $3k/(k+2)$.  This confirms that there are
not extra stable particles in the spectrum of $CP^1$.  In the general
$CP^{n-1}$ models, for example, the $\sigma$ and $\pi$ particles of \witN have
mass greater than twice the soliton mass (the mass $\rightarrow 2M$ as
$n\rightarrow\infty$), so they will decay into the solitons. In fact, there is
pole in $S_{1(n-1)}(\tht)$ corresponding to this supermultiplet of unstable
particles: it is at $\tht=2i\pi(\half + {1\over n})$, following from
\Pis\ with $\tht\rightarrow i\pi - \tht$.

\newsec{$N$=2 Sine-Gordon and Generalizations}

The $N$=2 supersymmetric sine-Gordon theory is described by the
action
\eqn\ssg{\int d^2zd^4\t X\bar X+(\int d^2zd^2\t \cos\beta
_{N=2}X+h.c.),}
where, as usual, $X$ and $\bar X$ are chiral and anti-chiral superfields.
While the parameter $\beta _{N=2}$ can be complex, the phase of $\beta _{N=2}$
can be eliminated by redefining $X$ and $\bar X$; we thus take $\beta _{N=2}$
real.  The theory \ssg\ is expected to be integrable as it is the
supersymmetric version of an integrable theory.  By our usual considerations,
the soliton content of \ssg\ consists of doublets $(u_p,d_p)$, with
$p\in {\bf Z}$, of equal mass connecting adjacent critical points of
the superpotential. The fermion numbers of $(u_p,d_p)$ are $(\half,-\half)$.

A natural conjecture \Kob\ for the S-matrix is simply a tensor product of the
$N$=0 sine-Gordon S-matrix (written out in ref. \rZandZ), governing scattering
in the soliton labels $p$, with the $N$=2 supermultiplet S-matrix \Smat , with
$n$=2, governing scattering of the $u$ and $d$:
\eqn\ssgsmat{S^{N=2}_{SG}(\beta _{N=2};\t )=S^{N=0}_{SG}(\beta
_{N=0}; \t
)\otimes S^{N=2}[n=2](\t).}
This is particularly sensible when one recalls that the S-matrices for the
$N$=0 minimal models with $\Phi _{1,3}$ perturbation can be found as
restrictions of $N$=0 sine-Gordon \rBLii, and the S-matrices in the analogous
$N$=2 models (with the least-relevant perturbation) are these restricted
sine-Gordon S-matrices tensored with $S^{N=2}[n=2](\tht)$ \refs{\rBL,\pk}. It
is natural to expect the tensor product structure to remain when the
restriction is removed. By studying the quantum-group symmetry, it was found
in \Kob\ that the couplings in \ssgsmat\ are related by
$$\beta^2_{N=2}={\beta^2_{N=0}\over 1-{\beta^2_{N=0}\over
8\pi}}.$$
This can be interpreted as $N$=2 non-renormalization: $\beta _{N=2}$=$\beta
_{N=0}^{bare}$.

{}From \CPSmat\ and \ssgsmat\ it follows that the S-matrix for supersymmetric
$CP^1$ equals that of the $N$=2 super sine-Gordon for
$\beta_{N=2}\rightarrow\infty$ \foot{This corresponds to $\beta_{N=0}^2=8\pi$,
the point at which the $N$=0 model has a Kosterlitz-Thouless phase transition
and at which the $U(1)$ symmetry group is enlarged to $SU(2)$. It is the
self-dual radius $R=1/\sqrt{2}$, in the normalization of \rGin. This phase
transition is where the sine-Gordon perturbation is of dimension (1,1); the
$SU(2)$ symmetry mixes it with the kinetic term. The perturbation is the state
with momentum {\it two} and winding zero, and is the operator-product of left
and right chiral SU(2) currents. This relation is often misunderstood; the
``KT'' transition referred to in \rGin\ comes from perturbing by the state
with momentum zero and winding {\it one}. The S-matrix becomes rational (no
trigonometric functions) at the $SU(2)$ point.  In spin-chain language, this
is where the XXZ model becomes the XXX model.}; thus, the theories are
``equal''.  We can eliminate $\beta _{N=2}$ from the superpotential and put it
into the kinetic term by a rescaling of $X$ and $\bar X$.  The results of
\rCV\ and \rCVii , therefore, do not depend on $\beta$.  The limit $\beta
_{N=2}\rightarrow \infty$ where $N$=2 super sine-Gordon equals supersymmetric
$CP^1$ then scales the kinetic term to zero, leaving only the superpotential.
In this light, it is not surprising that it was found in \rCVii\ that the
ground-state metrics for the two models are equal.

To obtain the TBA system of equations for \ssgsmat\ we need to find the TBA
system for the $N$=0 part.  We have obtained a TBA system of equations for the
$N$=0 sine-Gordon theory at arbitrary $\beta$ \un\ using the analysis of \rTS;
the TBA system of equations is rather ugly for generic values of $\beta ^2$.
However, the system greatly simplifies at the points $\beta_{N=2}^2=8\pi(k+1)$
for positive integer $k$. The use of the variable $k$ as in sect. 5 is not an
accident; it is at these values of $k$ that sine-Gordon can be restricted to
the $k$th Chebyshev minimal model.  We will, thus, limit our considerations to
these nice values of $\beta$.  The TBA system for the $N$=0 part of the theory
is of the usual form \TBA\ with species, masses, and $\phi _{a,b}$ described
by:

\bigskip
\noindent
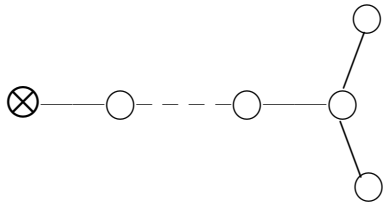
\centerline{\hbox{\rlap{\raise27pt\hbox{$\hskip4.3cm\bigcirc$}}
\rlap{\lower26pt\hbox{$\hskip4.2cm\bigcirc$}}
\rlap{\raise14pt\hbox{$\hskip3.95cm\Big/$}}
\rlap{\lower14pt\hbox{$\hskip3.8cm\Big\backslash$}}
{\raise1pt\hbox{$\bigotimes$}}------$\bigcirc$-- -- --
--$\bigcirc$------$\bigcirc$ }}
\bigskip

\noindent
where there are $k+1$ open nodes. The UV limit gives the central
charge $c=1$.

To obtain the TBA system for \ssg\ at these values of $\beta$, we tensor
product this $N$=0 TBA system with that for the $n=2$ case of our basic $N$=2
scattering theory, obtaining the diagram

\bigskip
\noindent
\centerline{
\hbox{\rlap{\raise28pt\hbox{$0\hskip.3cm\bigcirc\hskip
3.8cm\bigcirc$}}
\rlap{\lower27pt\hbox{$\bar 0\hskip.3cm\bigcirc\hskip
3.8cm\bigcirc$}}
\rlap{\raise14pt\hbox{$\hskip.55cm\Big\backslash\hskip3.6cm\Big/$
}}
\rlap{\lower14pt\hbox{$\hskip.5cm\Big/\hskip3.55cm\Big\backslash$
}}
$\hskip.45cm${\raise1pt\hbox{$\bigotimes$}}------$\bigcirc$-- --
--
--$\bigcirc$------$\bigcirc$ }}
\bigskip

\noindent The $N$=1 sine-Gordon model \refs{\rTsv,\rABL} at
analogous couplings, is, as in \pk , described by the above figure with the
$\bar 0$ node removed.  The reader can check that the UV casimir energy gives
the correct central charge $3/2$ for the $N$=1 case and $3$ for the $N$=2
case.  The $x_a$ are all $\infty$ in the $N$=2 case, as they are in all cases
described by an extended ADE Dynkin diagram.  As a result of $x_a=\infty$, the
leading UV corrections to the free energy depend on $\log mR$, and can be
calculated in the manner of $\rAlZiv$. These logarithmic terms also appear in
the $\beta _{N=2}\rightarrow \infty$ theory corresponding to the theory
$CP^{1}$ discussed in the previous section; analogous logarithmic
contributions were found in the $c$-function of \rCVii.

We can generalize \ssg\ to consider the $SU(n)$ affine Toda
theories described by the action
\eqn\ata{\int d^2zd^4\t \sum _{j=1}^{n-1}X_j\bar X_j+{1\over
2}(\int d^2zd^2\t
\sum _{j=1}^{n}e^{i\beta _{N=2}(X_j-X_{j-1})}+h.c.),}
where $X_0\equiv X_n\equiv 0$.  The theory has a \Zn\ symmetry corresponding
to $\exp (i\beta _{N=2}X_j)\rightarrow \omega ^j\exp (i\beta _{N=2}X_j)$ with
$\omega ^n=1$.  Our usual considerations reveal that the soliton masses and
fermion numbers are those of the basic \Zn\ theory \mrp .  We, again, expect
the theory to decompose as in \sprod, with the $N$=0 part the same as $N$=0
affine Toda theory at imaginary coupling \rHollo. The $N$=0 S-matrices should
be the \Zn\ generalizations of the sine-Gordon S-matrices discussed in \rBel.
In the lattice-model language, these are $SU(n)$ generalizations of the
six-vertex model. It then follows that the S-matrix of \ata\ at
$\beta_{N=2}\rightarrow \infty$ equals that of the supersymmetric
$CP^{n-1}$ sigma model, which is the $k\rightarrow \infty$ limit of the
theories of sect.\ 5.

\newsec{Conclusions}

While we have here analyzed a wide variety of integrable $N$=2 theories, there
is a simple class of integrable theories for which the scattering theory
remains unknown: perturbations of the $N$=2 minimal models corresponding to
the Landau-Ginzburg superpotential $W=X^n +X^2$ \refs{\toda,\rMW}. It is easy
to see that there is no completely elastic S-matrix involving only solitons;
there must be additional states \LGSol.  Presumably there are non-solitons in
the theory which can be created by soliton scattering. It is a peculiar
feature of many two-dimensional models that only solitons appear in the
spectrum; obviously there is no reason this has to be true for all integrable
models.  The constraints imposed by elasticity require that the non-solitons
have the same masses as the solitons. One example of a model with such a
structure is the sub-leading magnetic perturbation of the tricritical Ising
model, as discussed in \rAZtci.

There are many integrable $N$=2 models with much more intricate soliton
structures \rLW.  These theories presumably have $N$=0 analogs as do most of
the theories, and it would be amusing to find them.  One must, of course, be
careful in finding these analogs; as we have seen, the $N$=0 analogs of the
supersymmetric $CP^{n-1}$ models of sect.\ 6 are not the $N$=0 $CP^{n-1}$
models, but instead are affine Toda theories at a critical coupling.

In this paper and in \pk\ we have often discussed the results of \rCV .  The
results of \rCV\ provide an exact non-perturbative $c$-function for any $N$=2
theory via differential equations; integrability is not required, only
knowledge of the chiral ring.  The TBA provides an exact non-perturbative
$c$-function, proportional to the Casimir energy $E(R)$, via integral
equations; integrability and knowledge of the exact S-matrix and its transfer
matrix eigenvalues are required but $N$=2 supersymmetry is not.  The two
approaches are radically different and, by comparing on our common ground of
integrable $N$=2 theories, it is seen that the two $c$-functions are, in fact,
different.  We have, however, found a way to explicitly calculate the $c$
function of \rCV\ from a TBA calculation.  This remarkable link between the
results of \rCV\ and the TBA could provide a means to extend both approaches
into new realms.  Alas, this will have to wait for a future publication.
\bigskip
\centerline{\bf Acknowledgements}

We would like to thank R. Rohm and C. Vafa for valuable discussions.  We would
also like to thank C. Vafa for valuable arm-twisting. K.I.\ was supported by
NSF grant PHY-87-14654 and an NSF graduate fellowship, and P.F.\ was supported
by DOE grant DEAC02-89ER-40509.

\listrefs
\end